\documentclass[letterpaper,english,aps,reprint,citeautoscript,prb,twocolumn]{revtex4}
\usepackage[T1]{fontenc}
\usepackage[latin9]{inputenc}
\usepackage{amsmath}
\usepackage{amssymb}
\usepackage[english]{babel}
\usepackage{graphicx}

\newcommand{\vhat}[1]{\mathbf{\hat{#1}}}

\begin{document}

\title{Magneto-Optical Trap Field Characterization with the Directional Hanle Effect}

\author{Jarom S. Jackson, Dallin S. Durfee*}
\affiliation{Department of Physics and Astronomy, Brigham Young University, Provo, Utah 84602}

\begin{abstract}
We demonstrate the use of spatial emission patterns to measure the magnetic fields used in a magneto-optical trap.  The directional aspect of the Hanle effect gives a direct, visual presentation of the magnetic fields, in which brighter fluorescence indicates larger fields.  It can be used to determine the direction as well as the magnitude of the field.  It is particularly well suited for characterizing and aligning magneto-optical traps, requiring little additional equipment or setup beyond what is ordinarily used in a magneto-optical trap, and being most sensitive to fields of the size typically present in a magneto-optical trap.
\end{abstract}

\maketitle



\section*{Introduction}
In 1924, Wilhelm Hanle presented a theory explaining the depolarization of light scattered by an atomic vapor in a magnetic field. This effect, known as the Hanle effect, has been used to determine atomic state lifetimes \cite{moruzzi2013hanle}, to measure magnetic fields \cite{kastler73he,roux2008bose,budker02review}, and to measure phenomena related to magnetic fields \cite{Johnson85ic,OBrien16oa}. Other manifestations associated with the Hanle effect include level-crossing spectroscopy \cite{colegrove59levelcrossing,novikov74HanleEffect}, and a ``mechanical'' Hanle effect \cite{kaiser1991mechanical}.  Field measurements using the Hanle effect require no sensor other than the atoms themselves in the region being measured, and first-order Doppler shifts do not degrade the measurement precision \cite{novikov74HanleEffect}, such that precise measurements can be made with hot gasses. 

While most Hanle effect studies focus on depolarization, the interaction of atoms with magnetic fields also affects the spatial pattern of the scattered light \cite{WaltherAtomicBeam,ctmastereq,franken61crossedstates,novikov74HanleEffect}.   We will refer to the field-dependent spatial radiation pattern as the ``directional'' Hanle effect.  Unlike polarization measurements, directional Hanle effect measurements are sensitive to atom density and driving light intensity.  But they are not susceptible to errors due to polarization-altering properties of the medium between the measurement region and the detector, and they don't require polarizing elements in the detector or the need to rotate polarizers during the measurement.  Adding a polarizer can, however, reveal additional information.

Applications of the directional Hanle effect are possible in astrophysics, laboratory plasmas, and other disciplines where Hanle effect-related techniques have been developed \cite{stenflo1982hanle,stenflo1998hanle,bommier1980quantum} or proposed \cite{presura2012hanle,Ignace97he,Ignace10he}.  In this paper we demonstrate the utility of the directional Hanle effect to characterize the magnetic fields used in a magneto-optical trap (MOT).  While other fluorescence-based field measurements have been demonstrated \cite{courteille2001tomography, nielsen2008zeeman}, a significant advantage of the directional Hanle effect for MOT alignment and characterization is that it produces an immediate, visual display of the field strength, where brighter fluorescence corresponds to larger fields.  Simply looking at the fluorescing atoms, one can see where the field is large or small, and where field zeros are.  One can quickly identify the MOT trapping region as a dark spot in the fluorescence.  This visual feedback is particularly useful when aligning laser beams for a MOT.

The method can be applied with little or no additional equipment beyond what is already present in a typical MOT, usually requires little change to the MOT apparatus, and, as discussed later, is most sensitive to fields of the magnitude typically present in MOTs.  Because the effect saturates at about the same field magnitude which defines the edge of the trapping region, the size of the dark spot in the fluorescence gives an estimate of the size of the MOT trapping volume.  In addition to magnitude, the technique can also be used to determine the field's direction at every point in the measurement region.

\section*{Description of the Hanle effect}

The Hanle effect can be understood qualitatively by treating atoms as classical oscillating electric dipoles. In this picture, similar to what was first proposed by Hanle \cite{novikov74HanleEffect,budker02review}, light induces an oscillating electric dipole moment in an atom.  The dipole is initially excited in the direction of the driving light polarization.  If a magnetic field is present, the dipole oscillation axis will rotate around the field, changing the polarization of the emitted light.  Averaged over time, this reduces the scattered light polarization.  This is typically observed using a detector displaced from the atoms perpendicular to the driving light polarization, where the scattered intensity is greatest. If the field is large enough to cause many rotations within a decay time (which happens when Zeeman shifts are large compared to the natural linewidth), the light scattered in this direction will be completely unpolarized.

\begin{figure}
  \includegraphics[width=8 cm]{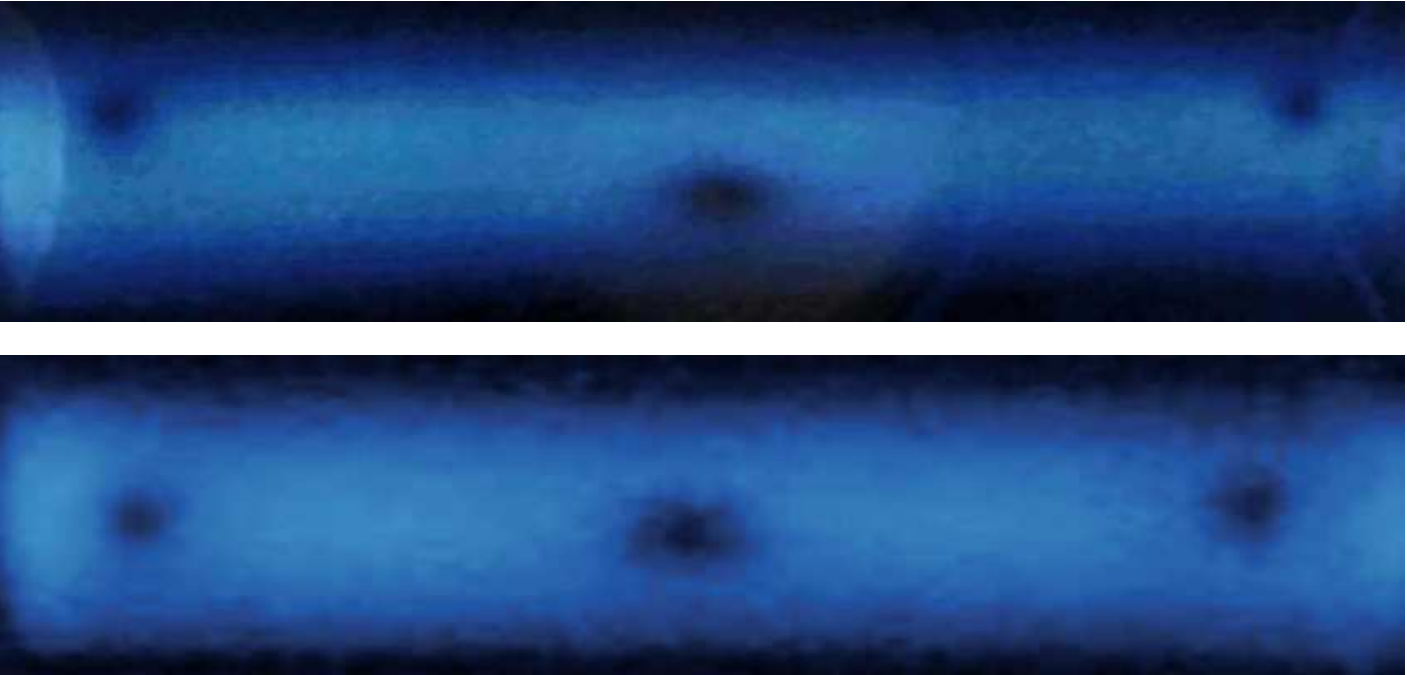}
  \caption{Two pictures of fluorescence from strontium atoms in an inhomogeneous magnetic field.  An expanding thermal strontium beam was illuminated with a thin sheet of resonant laser light.  With the laser polarization set along the line from the atoms to the camera, zero crossings in the field are clearly visible as dark spots. The width of these dark spots is related to the local field gradients.  The central zero point is the center of the MOT's trapping volume.  The other two dark spots are the points at which the laser light sheet intersects with a ring-shaped zero crossing caused by the non-traditional field geometry in our apparatus.  The fact that the zero crossings are far from being collinear in the upper figure revealed to us that the magnets generating the field had significantly different magnetizations.  The lower image was taken after the defective magnet was replaced.
}
	\label{fig:3ZeroPoints}
\end{figure}

In the {\em directional} Hanle effect, we consider how the field changes the {\em spatial radiation pattern} of the scattered light.  With no field present, the oscillating dipole emits with a sine-squared dipole emission pattern, and a detector displaced from the atoms in the driving light polarization direction will measure no scattered light.  If a magnetic field is present, however, the emission pattern rotates.  The strength of the field determines the rotation rate, and the rotation rate, compared to the decay time, determines the average light intensity scattered in the direction of the pump light polarization.  An example of this can be seen in Fig.~\ref{fig:3ZeroPoints}, which shows fluorescence from atoms in a region with 3 field zero points \cite{jackson2016situ}.  Note that the directional Hanle effect provides a picture of the field strength, viewable with a camera or directly viewable by eye.

\section*{Measurement of MOT fields}

\begin{figure}
  \includegraphics[width=8cm]{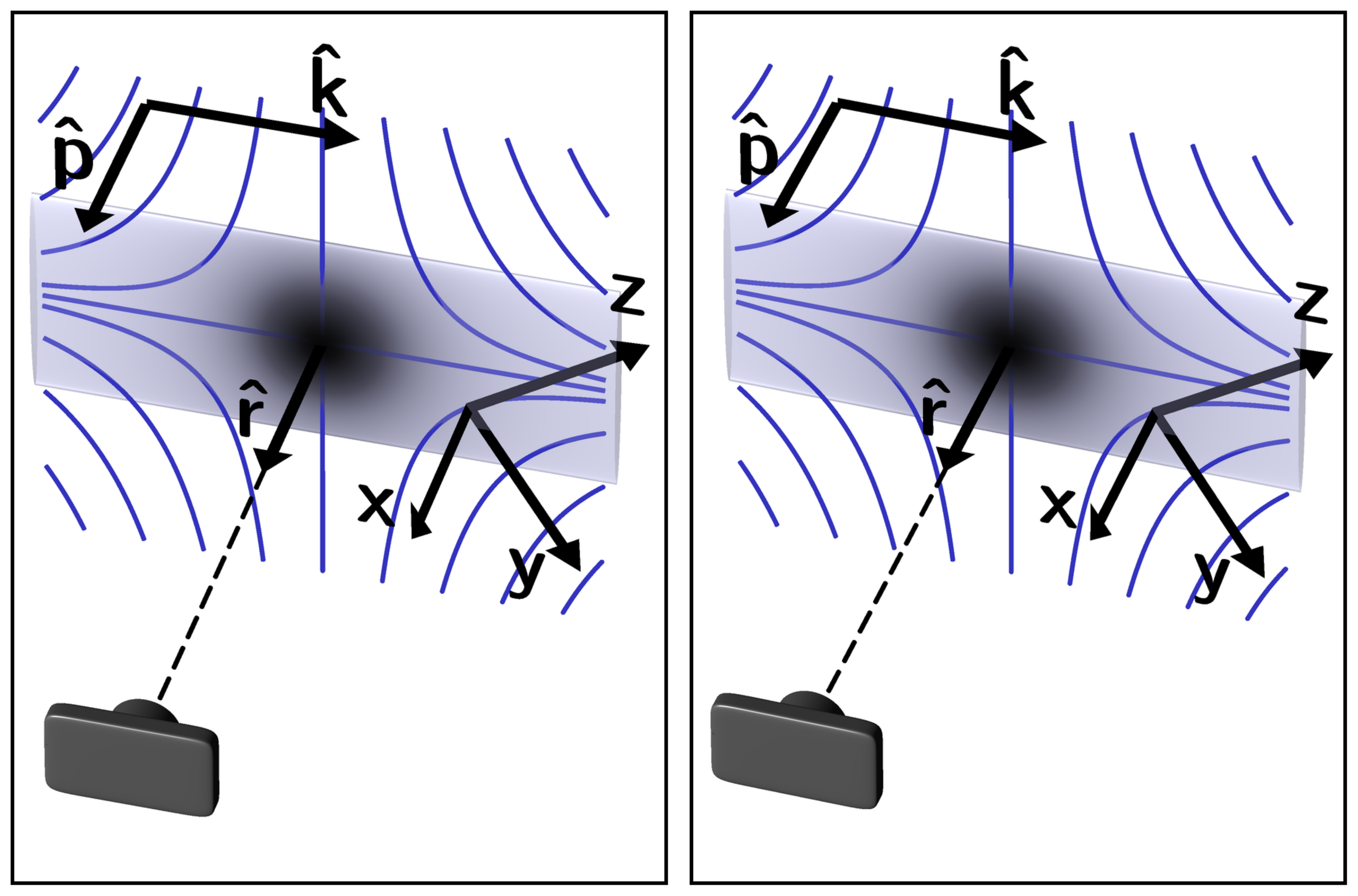}
  \caption{Illustration of the coordinate system.  The two images were rendered from a slightly different perspective to produce a 3D stereoscopic pair for divergent `wall-eyed' viewing\cite{viewingstereopairs}.  The sheet of light, polarized in the $\vhat{p}$ direction, is depicted by a rectangle.  The dark area in the center of the rectangle depicts a low-field region from which little light scatters toward the camera, shown in the lower left-hand corner displaced from the center of the rectangle in the $\vhat{r}$ direction.  The curves represent magnetic field lines.  The field direction defines the local $z$ axis, and $x$ is defined to be in the $\vhat{p}$ direction, which is orthogonal to the laser propagation direction $\vhat{k}$. }
  \label{fig:schematic}
\end{figure}

We used the directional Hanle effect to characterize the inhomogeneous magnetic field in the center of a strontium MOT apparatus  \cite{erickson2011development}.  To do this, we blocked all but one laser beam going through the vacuum chamber.  With just one beam there are no cooled, trapped atoms.  Instead, thermal atoms from the oven used to load the trap scatter light from the remaining beam.  A narrow slit was placed in the beam to create a sheet of light passing through the center of the MOT trapping region.  The light sheet was aligned to pass through the field zero in the center of the region by adjusting its position to make the central dark spot in the image as large and dark as possible.  The scattered light was measured with a camera displaced from the atoms in a direction $\hat{r}$ which is parallel to $\hat{p}$, as shown in Fig.\ \ref{fig:schematic}.

To analyze the data in this paper, we used a quantum model derived using a master equation\cite{ctmastereq}.  For quantitative measurements, weak driving light, such that the peak saturation parameter is much less than one, is preferred.  Otherwise the pump light intensity at each point in the image needs to be known with some precision.  We assume that the probe field is not large enough for inelastic scattering, and we assume that the atomic vapor is dilute, such that re-scattering of scattered light is negligible.  Because of the known symmetry of the field, we assume that the field at every point in the light sheet lies within the plane of the sheet.

For a $j=0$ to $j=1$ transition driven with linearly polarized narrow-band radiation, in the limit of weak pump light, and assuming that the magnetic field is orthogonal to $\hat{p}$, the intensity scattered in the $\hat{p}$ direction is proportional to
\begin{equation}
   I_{\hat{p}} = \frac{\omega_L^2 }{ \Delta \omega^4+2 \Delta \omega^2 \left(\left(\frac{\Gamma}{2}\right)^2-\omega_L^2\right)+\left(\left(\frac{\Gamma}{2}\right)^2+\omega_L^2\right)^2}
\label{eq:Ip}
\end{equation}
where $\Delta \omega$ is the detuning from resonance (including both laser detuning and Doppler shift), $\Gamma$ is the natural linewidth of the transition, and the Larmor frequency $\omega_L = g \mu_B B/\hbar$, where $\hbar$ is Planck's constant over $2\pi$, $\mu_B$ is the Bohr magneton, $B$ is the field magnitude, and $g$ is the Land\'e g-factor ($g=1$ for the 5s5p $^1$P$_1$ state in strontium).   For comparison, the light that would be scattered in a direction {\em orthogonal} to $\hat{p}$, in the {\em absence} of a magnetic field is proportional to
\begin{equation}
  I_{perp} = \frac{1}{\Delta \omega^2+\left(\Gamma/2\right)^2}
\label{eq:Iperp}
\end{equation}
with the same constant of proportionality.

The 5s$^2$ $^1$S$_0$ to 5s5p $^1$P$_1$ resonance transition in strontium has a simple structure, such that we don't have to consider effects such as the ground-state Hanle effect \cite{Arimondo01HanleClosed,Singh15GroundStateHanleInRb}, ground-state coherences and dark resonances \cite{arimondo07DarkResonances}, and optical pumping \cite{novikov74HanleEffect}.  In systems where optical pumping occurs, the same repump light used to realize the MOT can be employed when taking directional Hanle effect measurements.  

\section*{A simplified model}

In addition to the full model we used in our analysis, an extremely simplified model of light scattering can be useful to develop intuition.  While this model is not as accurate as the one we used to analyze data, it is easy to visualize the derivation, and it results in a simple equation from which insight into the directional Hanle effect can be easily gained.

The Hanle effect can be thought of in terms of quantum interference and the averaging of quantum beats \cite{budker02review,novikov74HanleEffect}.  In the simplified model we calculate the beating contributions from different emitting states to the light emitted in the direction of the driving light polarization.  This simplified model is a semi-classical model in which we treat radiation fields classically, but quantize atomic energy levels.  To generate this simple model, instead of atoms which are continuously driven, we consider an atom which is excited by an impulse.  After absorbing a photon, an atom's initial state will be a superposition of upper-state magnetic sublevels.  Each sublevel emits a classical field, and these fields interfere.

As with our more complete analysis, we will assume a $j=0$ to $j=1$ transition.  This simple level structure makes the mathematics less complicated and more intuitive, and avoids the need to consider things such as dark states and optical pumping.  For this level structure, our treatment is in qualitative agreement with the classical oscillating dipole description given above.  Furthermore, this is the structure of the transition used in our experiments.  

Because the field in our measurement region lies within the plane of the light sheet, orthogonal to $\hat{p}$, the $m=0$ upper state is not excited.  In this simplified model we will assume Zeeman shifts only affect the relative phase evolution of the upper states, not the overall scattering rate, and further assume that the $m=-1$ and the $m=+1$ states are populated equally, ignoring effects of laser detuning and Doppler shifts, which would result in different upper-state populations.  For broadband driving light, this is approximately true.  With narrowband excitation, when the Doppler width of the atoms is much larger than both the natural linewidth and the Zeeman shifts, errors due to this approximation largely wash out when emissions are averaged over the velocity distribution of the atoms.  Even when it is not a good approximation, it allows us to find a simple analytical expression to aid intuition. 

For the simplified model we assume that each of the two excited states emits the field of a classical rotating dipole (rotating in opposite directions), which decays according to the transition linewidth $\Gamma$.  Due to Zeeman shifts, the components of the field from the two sublevels oscillate at slightly different rates, $\omega_{\pm} = \omega_0 \pm \Delta \omega$, where $\omega_0$ is the unshifted frequency, and $\Delta \omega = \mu_B g B/\hbar$.  This causes the relative phases of the two field contributions to drift, altering the radiation field in a way analogous to the field of a classical oscillating dipole with a drifting oscillation axis.

In the far field, the resulting time-averaged intensity is an inverted Lorentzian:

\begin{equation}
  \label{eq:Iavg}
  \left<I_\text{ensemble}\right> =  \frac{I_{perp}}{2} \left[1-\frac{1}{1 + \frac{4 \mu_B^2 g^2 }{\hbar^2 \Gamma^2}B^2} \right],
\end{equation}
where $I_{perp}$ depends on the driving light intensity and the density of atoms, and is equal to the intensity which would be detected with no field if the detector were displaced in a direction perpendicular to $\hat{p}$.

Eq.\ \ref{eq:Iavg} indicates, as expected, that in the limit of zero magnetic field, no light is emitted in the $\hat{p}$ direction.  As the field increases, the intensity grows, approaching an asymptotic limit of $I_{perp}/2$. The half-width, half-maximum width of the Lorentzian is
\begin{equation}
  \label{eq:lorentzianWidth}
  B_\text{HWHM} = \hbar \Gamma/2 \mu_B g.
\end{equation}
This is the field at which the Larmor frequency is half the natural linewidth.  Since the slope of Eq.\ \ref{eq:Iavg} flattens once $B$ is several times $B_\text{HWHM}$, and because optical trapping is inefficient when the Zeeman shift becomes comparable to or larger than the light detuning, which is typically on the order of the natural linewidth in a MOT, the size of the dark region in the fluorescence is related to the size of the MOT trapping volume.  Because of the shallow slope of the curve near zero field and the asymptotic saturation at high fields, measurements using the directional Hanle effect are most precise for field magnitudes of order $B_\text{HWHM}$.  This is precisely the scale of the fields present in a typical MOT's trapping region \cite{budkeratomicphysics}.

\begin{figure}
  \includegraphics[width=8cm]{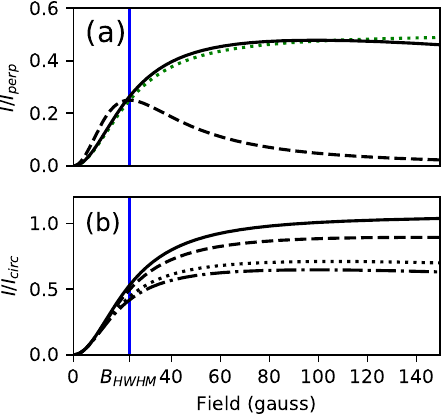}
  \caption{Intensity emitted along the pump light polarization axis vs.\ field strength. In (a) the theoretically expected light intensity scattered in the $\hat{p}$ direction, normalized to $I_{perp}$, is plotted as a function of magnetic field.  The full model, applied to motionless atoms with the driving light tuned to resonance, is shown with a dashed line.  The solid line represents the full model applied to a thermal beam of atoms, as described in the text.  The dotted line is the simplified model.  In (b) the full model is shown  normalized to the average of $I_{LH}$ and $I_{RH}$.  All curves assume that the field direction $\hat{z}$ is in the plane of the light sheet.  The solid/dashed/dotted/dash-dotted lines in (b) represent magnetic fields at an angle of 0/30/60/90 degrees from the $\hat{k}$ direction.  In both (a) and (b) the blue vertical line represents $B_{HWHM}$.}
          \label{fig:lightcurves}
\end{figure}

\section*{Theoretical curves}

The expression in Eq.~\ref{eq:Iavg} is graphed in Fig.~\ref{fig:lightcurves}(a). Also shown are curves generated using the full model of Eq.\ \ref{eq:Ip} which we used to analyze data.  In the simplified model, laser intensity, detuning, and Doppler shifts only impact $I_{perp}$, not the overall shape of the curve.  In the full model these factors are more important.  We assumed a thermal strontium beam from an oven with a temperature of $509^\circ$~C propagating in the opposite direction from the driving light, which was tuned to cancel the Doppler shift at the peak of the velocity distribution.  For the full model curves, $I_{\hat{p}}$ and $I_{perp}$ were first averaged over the atomic thermal distribution, and then the two averages were divided to generate the curves in Fig.\ \ref{fig:lightcurves}(a).

Fig.~\ref{fig:lightcurves}(a) shows that for a gas of thermal atoms with Zeeman shifts that are small compared to the Doppler broadening, the simplified solution is a good representation of the effect, deviating from the full model by only a few percent.  While many treatments of the Hanle effect using broadband excitation result in similar Lorentzian curves \cite{moruzzi2013hanle}, this figure illustrates that Lorentzian curves are also useful under narrowband excitation as long as the Doppler broadening is large.  The errors introduced into our simplified model by assuming an impulse drive and constant scattering rates mostly vanish when both models are normalized to $I_{perp}$, as $I_{perp}$ includes the same assumptions.  

For hot atoms, the scattered intensity in the full model initially increases somewhat more rapidly with field than the simplified model.  Atoms whose Doppler shift makes them not exactly resonant with the pump light will have one upper state Zeeman shifted {\em closer} and the other {\em further} from resonance, such that one state scatters more than the other.  This makes complete destructive interference impossible.  At high fields, such that the magnitude of the Zeeman shift surpasses both the natural and the Doppler linewidth, the bulk of the atoms in the distribution get shifted from resonance, and the intensity in the full model asymptotically approaches zero with increasing field.  While readily visible in Fig.\ \ref{fig:lightcurves}(a) for cold atoms, for the thermal distribution of atoms in our experiment, this effect only becomes significant at fields higher than those shown in Fig.~\ref{fig:lightcurves}.

\section*{Calibration}

Both the simplified and complete model require some knowledge related to the light intensity and atomic density at each location in the image.  In the limit of low drive intensity (such that the saturation parameter $s \ll 1$), only the product of the two must be known.  In this limit, the intensity $I_{perp}$ is proportional to this product.  So a simple way to calibrate field measurements, without needing to know the atom density or light intensity at any point in the measurement region, is to zero the magnetic field, rotate the driving light polarization such that it is orthogonal to $\hat{r}$, and directly measure $I_{perp}$.  In our setup, we were unable to easily remove the magnets, and were not able to make this measurement.  

Another approach is to make assumptions about the spatial distribution of the driving light and the atomic vapor, and about the field at certain locations.  For example, we achieved results consistent with the expected field by assuming a constant vapor density and assuming that the light scattered along vertical slices half-way between the zeros in Fig.\ \ref{fig:3ZeroPoints} was near the maximum of the curve in Fig.\ \ref{fig:lightcurves}(a).  While the field is not constant along these slices, with these assumptions, and assuming a collimated laser beam, we obtained results which fit the expected shape of the field quite well.  However, the overall magnitude of the field and the size of the field gradients at the location of the central zero were very dependent on where on the curve of Fig.\ \ref{fig:lightcurves}(a) we assumed the vertical slices to be.

A better approach, when the field can't be zeroed, is to measure the light scattered by circularly polarized driving fields.  The amount of light scattered in the $\hat{r}$ direction is less sensitive to the field when circularly polarized pump light is used (see Fig.\ \ref{fig:polimages}(f)).  Simply dividing by an image made with circularly polarized light and using the simplified model yields a reasonable result.  To obtain more accurate results, we extended the full model of the Hanle effect\cite{ctmastereq} to allow circularly polarized driving light.  In the limit of low pump intensity, and assuming the magnetic field lies within the plane of the light sheet, the light scattered to the camera by a circularly polarized pump beam is proportional to
\begin{equation}
\begin{split}
I_{LH/RH} & 
=   \frac{I_{\hat{p}}}{2 \omega_L^2} \biggl( 
\left(\Delta \omega^2+(\Gamma/2)^2\right) \cos ^2(\theta) \\
&\quad 
\pm 2 \Delta \omega \omega_L \cos (\theta)+\omega_L^2 \biggr) \\
&\quad 
+ \frac{I_{perp} }{2} \sin ^2 (\theta)
\end{split}
\label{eq:Icirc}
\end{equation}
where $\theta$ is the angle of the magnetic field relative to $\hat{k}$, and the upper/lower sign is used to find the intensity scattered by left-/right-handed polarized light.  If the same intensity of pump light is used, the proportionality constant is the same for Eqs.\ \ref{eq:Ip}, \ref{eq:Iperp}, and \ref{eq:Icirc}.  Otherwise, these equations must be scaled accordingly.  Normalizing $I_{\hat{p}}$ to $I_{circ} = (I_{LH} + I_{RH})/2$ results in the curves shown in Fig.\ \ref{fig:lightcurves}(b).

\section*{Results}
\label{sec:results}

\begin{figure}
  \includegraphics[width=8cm]{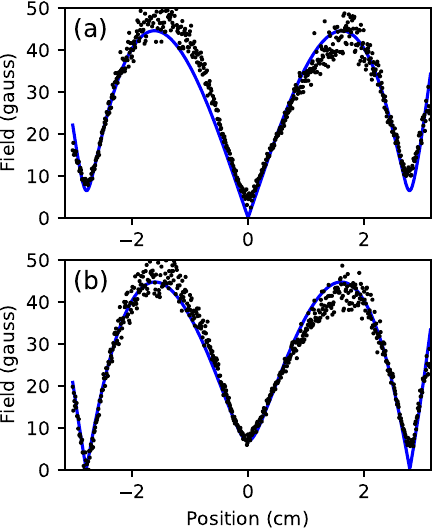}
  \caption{Magnetic field measurement.  Using the full model normalized to data taken using circularly polarized light, the magnetic field along two horizontal lines of an image similar to the lower image in Fig.~\ref{fig:3ZeroPoints} was determined.  The measured field is shown as dots.  The solid line is the expected field for a pair of opposing ring magnets. Plot (a) shows the field along a horizontal line passing through the center of the central field zero.  In (b) a similar plot is shown for a horizontal row of data displaced 0.85 mm vertically from the data used for the plot shown in (a).}
\label{fig:slice}
\end{figure}

The magnetic field measured along two horizontal lines are shown in Fig.~\ref{fig:slice}, along with the field we predicted using the known geometry of the magnets generating the field.  The only free parameters used to fit the model to our measurements were the size of the effective surface currents (the $BHc$ coercive field force) of the two magnets, and a 4\% adjustment to the effective size we had estimated for the pixels in the images.  Note that for fields much larger than $B_\text{HWHM}$, small changes in intensity correspond to large changes in magnetic field, such that small amounts of noise result in large errors in the measured field.  This accounts for increased noise in regions of Fig.\ \ref{fig:slice} with larger fields.

The outer two minima in Fig.\ \ref{fig:slice}(a) don't go to zero because of a known slight mismatch in the $BHc$ of the two ring magnets generating the field.  This offsets the ring-shaped zero crossing axially from the central zero crossing.  Fig.\ \ref{fig:slice}(b) shows the predicted and measured field magnitude along a horizontal line displaced 0.85 mm vertically from the center of the central field minimum.  At this vertical displacement, both the predicted and measured field drop close to zero at the location of the outer field minima.

While the data fits the expected field shape extremely well, in order to make the predicted field match our measurements, we had to set $BHc$ for the two magnets to 1191 and 1217 kA/m, slightly above the expected range of 860-955 kA/m for N42 neodymium magnets \cite{magnetstrengths}.  The 2\% difference in the magnetization of the two magnets agrees with Hall probe measurements of the magnets.  The horizontal field gradient at the central zero, as determined using the directional Hanle effect, is 41 gauss/cm.  This is somewhat larger than the $30 \pm 4$ gauss/cm that we measured with a Hall probe with the magnets mounted in a test assembly.  We are not certain why our measurements are slightly larger than expected.  

\section*{Field direction}
\label{sec:FieldDir}

With a slight extension to this method, the local {\em direction} of the magnetic field can also be determined.  This is done by first measuring the polarization of the scattered light emitted in the $\hat{p}$ direction, and then noting the asymmetry in the light scattered at low fields when viewed from a direction at an angle to the $\hat{p}$ direction.

\begin{figure}
  \includegraphics[width=8cm]{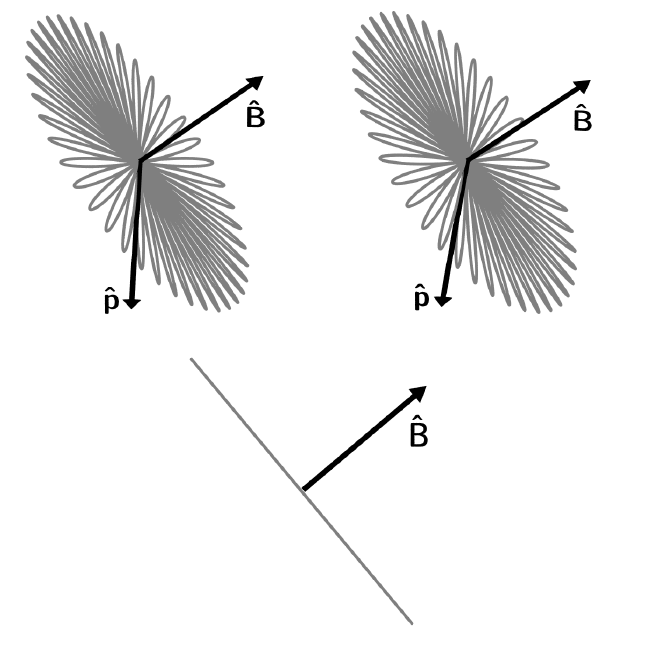}
  \caption{The polarization of the scattered light.  The direction of the driving light polarization is indicated with the arrow labeled $\hat{p}$.  The magnetic field direction is shown with the arrow labeled $\hat{B}$.  A ``cartoon'' illustration of the path that the oscillating dipole moment takes is shown in gray.  In the upper image pair (a stereo pair for divergent `wall-eyed' viewing), we see that the oscillation direction rotates in a plane orthogonal to the field direction.  In the bottom image, we see the same path as viewed from a point along the line defined by $\hat{p}$.  From this vantage point, the projection of the oscillating dipole moment we see, and therefore the polarization of the light we measure, is always orthogonal to the field direction.
}
	\label{fig:linpol}
\end{figure}

\begin{figure}
  \includegraphics[width=8 cm]{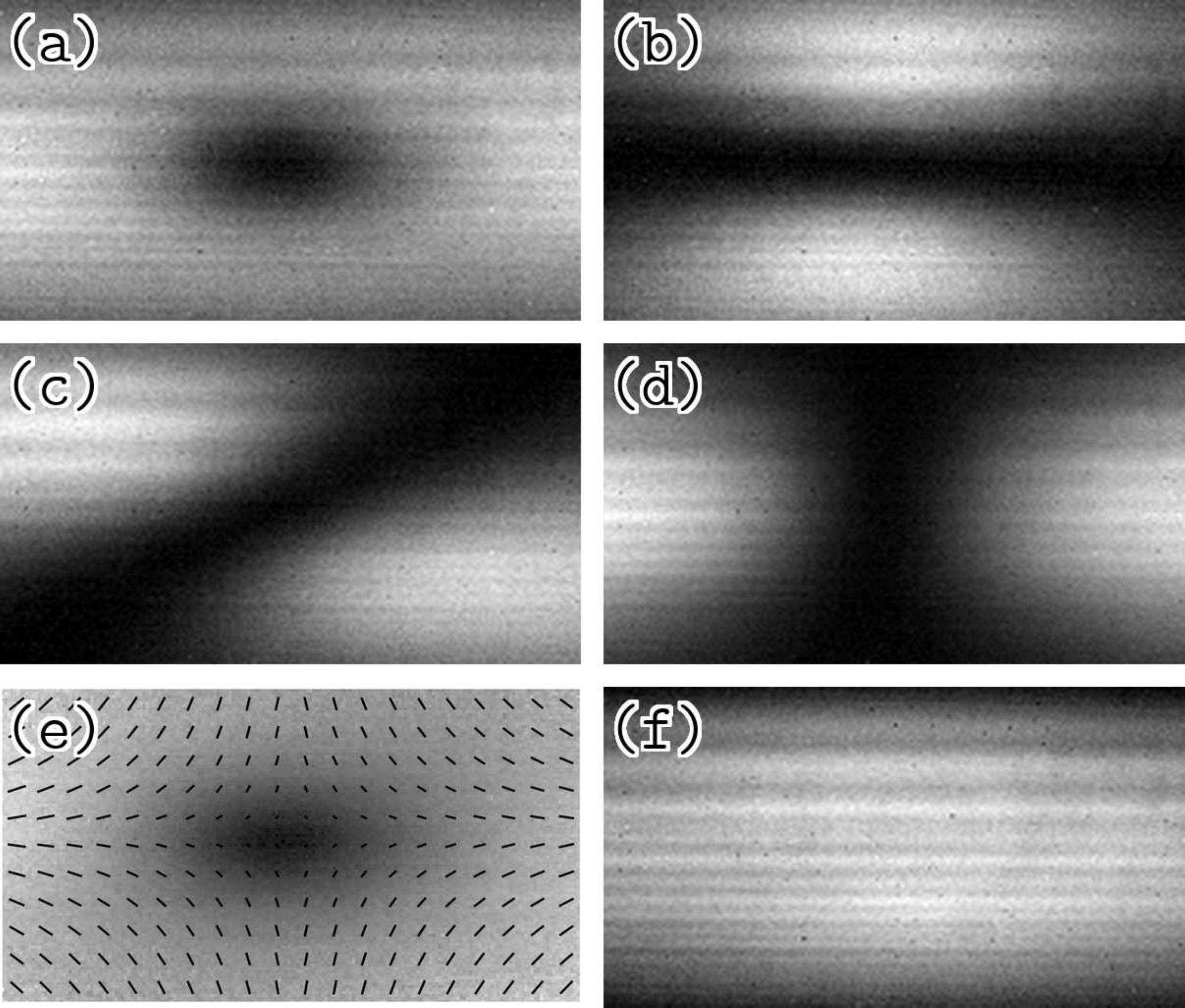}
  \caption{Images taken through a polarizer.  Images of the fluorescing atoms were taken without a polarizer (a), and through a polarizer which allowed light to pass at a polarization angle of $0^\circ$ (b), $45^\circ$ (c), and $90^\circ$ (d) from horizontal.  The vector graph in (e) shows the direction of the field (to within an overall sign) calculated from the above images. Image (f) shows the average of two pictures taken with left- and right-handed circularly polarized light, in both cases without a polarizer in front of the camera.  The horizontal stripes in the images are due to variations in pump light intensity across the light sheet.
}
	\label{fig:polimages}
\end{figure}

If, as previously assumed, the magnetic field direction is perpendicular to $\hat{p}$, the light scattered in the $\hat{p}$ direction will be linearly polarized orthogonal to the magnetic field.  We can understand this intuitively by thinking of the oscillating atom as a classical oscillating dipole.  The oscillation axis, initially excited along the $\hat{p}$ direction, will rotate around the magnetic field, as illustrated in Fig.~\ref{fig:linpol}.  Figs.~\ref{fig:polimages}(b-d) show images of fluorescence measured with a linear polarizer in front of the camera with its transmission axis at various angles.

By taking three pictures with a linear polarizer placed in front of the camera --- one with the polarizer aligned to pass light polarized in the horizontal direction, one with the polarizer rotated at 45 degrees from the horizontal, and one aligned to pass vertically polarized light, the field direction can be found.  
If $\hat{B}$ is a unit vector in the direction of the field at a point in the images, the magnitudes of the components of $\hat{B}$ are
\begin{equation}
  |\hat{B}_h| = \sqrt{\frac{I_v}{I_h + I_v}} \quad \mbox{and} \quad
  |\hat{B}_v| = \sqrt{\frac{I_h}{I_h + I_v}}
\end{equation}
where $I_h$ and $I_v$ are the intensities at this point in the images taken using a horizontal and vertical polarizer, respectively.  This narrows the direction of the field down to four possibilities: $\hat{B} = $ $\pm|\hat{B}_h|\hat{h}$ $\pm |\hat{B}_v|\hat{v}$, where $\hat{h}$ and $\hat{v}$ are unit vectors in the horizontal and vertical directions.  Using the image in Fig.~\ref{fig:polimages}(c), we can reduce this to two possibilities.  The light intensity at this point in this image, $I_d$, is related to the component of the magnetic field in the $(\hat{h} + \hat{v})/\sqrt{2}$ direction:
\begin{equation}
  \frac{1}{\sqrt{2}} \left| \hat{B}_h + \hat{B}_v \right| = \sqrt{\frac{I_d}{I_h+I_v}}
\end{equation}
From this, one can show that
\begin{equation}
  \hat{B}_h \hat{B}_v = \frac{I_d}{I_h+I_v} - \frac{1}{2}.
\end{equation}
Calculating this we can determine the sign of $\hat{B}_h \hat{B}_v$ at each point in the image, which then tells us if the signs of the vertical and horizontal components of $\hat{B}$ are the same or different.  This gives us the direction of the magnetic field at this point, to within an overall sign.  The results are shown in Fig.~\ref{fig:polimages}(e).  Note that if the orthogonal $45^\circ$ polarization axis is used, one side of the above equation must be multiplied by $-1$.  

\begin{figure}[h]
  \includegraphics[width=8 cm]{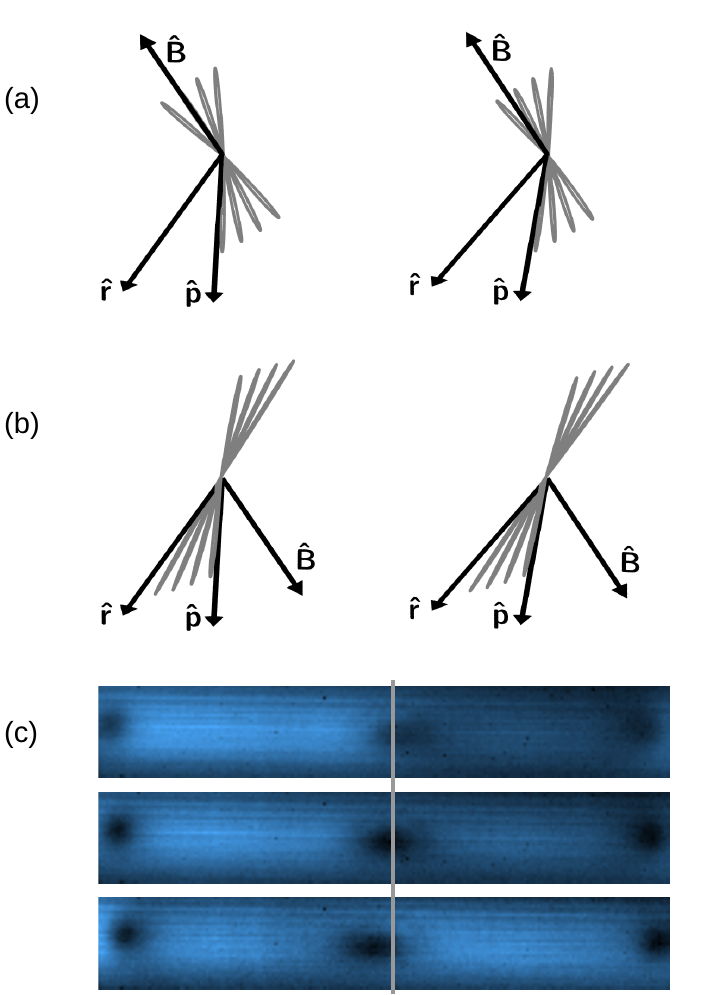}
  \caption{Scattered light asymmetry.  The camera in illustrations (a) and (b) is offset from the image region in the $\hat{r}$ direction, which is different from the $\hat{p}$ driving light polarization direction.  When the oscillating dipole is initially excited in the $\hat{p}$ direction, there is a non-zero projection of the oscillation as seen by the camera.  As shown in (a) (a stereo pair for divergent `wall-eyed' viewing), the magnetic field could potentially rotate the dipole oscillation direction further away from $\hat{r}$, increasing the light scattered to the camera.  As shown in (b), if the field is in the other direction, the oscillation direction rotates closer to $\hat{r}$, decreasing the light scattered to the camera.  The images in (c) were taken with the $\hat{p}$ polarization direction rotated at different angles relative to the $\hat{r}$ direction.  The middle image was taken using light polarized in the $\hat{r}$ direction.  The upper/lower image was taken with the light polarization rotated $20^\circ$ upward/downward from $\hat{r}$.  The gray line marks the horizontal center of the central zero in the middle image.  With the polarization rotated upward from $\hat{r}$, light levels decrease relative to the same point in the $\hat{p} = \hat{r}$ picture if the field is pointed to the right, and increase if it is pointed to the left.  With the polarization rotated downward, light levels decrease relative to the $\hat{p} = \hat{r}$ image if the field points to the left, and increase if it points to the right.
}
  \label{fig:offaxisasymmetry}
\end{figure}

To determine the sign of the field at certain locations, we can take an image without a polarizer in front of the camera, but with the camera location offset from the $\hat{p}$ direction.  This can be done by moving the camera or rotating the polarization of the driving light.  When this is done, the projection of the initial dipole oscillation direction onto the plane of the camera will have a non-zero length.  So even with no magnetic field some light will scatter to the camera.  In regions of low field, such that the oscillation axis of the dipole only rotates a small amount before the oscillation has decayed, the magnetic field can cause the light intensity to go up by causing the dipole oscillation axis to rotate further away from the camera's line of sight.  But it can also make it go {\em down} by rotating the dipole oscillation axis {\em toward} the line of sight, as illustrated in Figs.~\ref{fig:offaxisasymmetry}(a-b).

\onecolumngrid

\begin{figure}
  \includegraphics[width= 6.5 in]{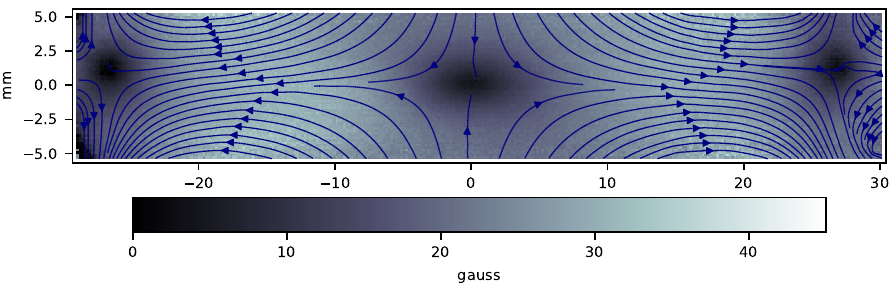}
  \caption{Magnetic field magnitude and direction.  The direction of the magnetic field at different points in the image, determined using the images in Fig.~\ref{fig:polimages} and  Fig.~\ref{fig:offaxisasymmetry}(c), is shown, along with the magnitude of the field at each point.
}
	\label{fig:fielddir}
\end{figure}

\twocolumngrid

Images of the fluorescence, taken with the $\hat{p}$ axis rotated relative to $\hat{r}$, are shown in Fig.~\ref{fig:offaxisasymmetry}(c).  One can infer the sign of the vector field shown in Fig.~\ref{fig:polimages}(e) from the asymmetry in these images.  Because the dark region in the top image in Fig.~\ref{fig:offaxisasymmetry}(c) is displaced to the right, when compared to the middle image, we know that the field to the right of the dark region is in the $+k$ direction, such that it rotates the oscillating dipole axis {\em toward} the camera, {\em reducing} the amount of light detected, and that the field at the left of the dark region is in the $-k$ direction, such that it rotates the oscillating dipole axis {\em away} from the camera, {\em increasing} the amount of scattered light detected.  Assuming that the field direction changes smoothly on the scale of a pixel, the direction of the field at other points can be inferred.  Combining the information from this image with our other measurements, we can unambiguously determine both the magnitude and the direction of the field at each point in the image.  The results are shown in Fig.~\ref{fig:fielddir}.

\section*{Discussion}
In conclusion, we demonstrated the characterization of magnetic fields used in a MOT using the directionality of scattered light in the Hanle effect. The method gives a direct visualization of the fields (including the location of field zero crossings and an estimate of the size of the MOT trapping volume), is most sensitive to field strengths typical of those found in MOTs, requires little additional equipment or setup, and does not require a physical probe in the vacuum (other than the atoms themselves), making it useful for MOT alignment and field characterization.  We used this technique to measure the magnitude and direction of the magnetic field at each point in a region of a plane passing through the center of a strontium MOT's trapping volume.  We also discussed methods to calibrate these measurements without the need to know the atomic density or pump light intensity at any point in the measurement region.

\section*{Data availability}
The datasets generated during and/or analyzed during the described study are available from the corresponding author on reasonable request.

\section*{Methods}

Most of the details of our experimental methods have already been discussed in the main text.  We point out here some additional details about our apparatus and methods used to collect and analyze data.

The magnetic fields in our MOT are generated by a pair of N42 neodymium ring magnets.  The rings have an inner/outer diameter of 2 and 3 inches, respectively, and a 1/2 inch axial thickness.  They are displaced from each other by 2.96 inches in the axial direction, and they are oriented such that their fields cancel at a point midway between them.  One can think of a ring magnet as a thin sheet of current running along the inner surface of the ring, and an opposing sheet of current running along the ring's outer diameter, such that each ring magnet is equivalent to two electromagnet coils with currents of equal magnitude running in opposite directions, such that their fields would cancel if not for the difference between the two coil diameters.  

The two magnets together make a field approximating a spherical quadrupole field near the midpoint between the magnets along the symmetry axis.  A ring-shaped zero crossing can occur, depending on the dimensions and separation of the rings, displaced radially from the central field zero.  Early field measurements showed an asymmetry, probably due to a magnet that had been weakened when accidentally overheated (see the upper image in Fig.~\ref{fig:3ZeroPoints}).  The data analyzed in this paper was taken after this magnet was replaced to create a significantly less asymmetric field (see the lower image in the same figure).

It is not important which beam is used to illuminate the atoms.  When using this technique to align MOT laser beams to the central field zero crossing, only the beam currently being aligned should be incident on the atoms.  For the data presented and analyzed in this paper, the unblocked beam was a 461 nm laser beam used to slow atoms entering the trap from an oven.  Because the light in this beam was initially circularly polarized, a linear polarizer was placed in the beam, and the polarization adjusted such that it was in a direction $\hat{p}$ which was normal to the surface of the light sheet.  

The camera we used to take our data was an inexpensive Logitech QuickCam Pro 9000 webcam with internal processing disabled \cite{richards2013logitech, langer2013logitech}.  With internal processing disabled, the pixel values are proportional to the intensity at each pixel, and we have access to the individual pixels of each color.  Because the 461 nm laser light used in the experiment mainly registers on the blue pixels, we selected only those pixels for our analysis.  

The driving light in our experiments had a peak saturation parameter of about 0.09 in the center of the pump beam.  With the camera we used, it was difficult to collect good data with a weaker beam.  For our analysis, we assumed the low saturation limit.  This avoided the problem of needing to know how the intensity of driving light changed across the laser beam profile.  Using a model that includes saturation effects, we determined that along the center of the image, where the intensity is the brightest and errors from this assumption are the largest, this would produce errors in the measured magnetic field on the order of 10 percent.  Incidentally, optical saturation effects are too small and in the wrong direction to account for the larger than expected magnetization fit parameters and field gradients discussed in the results section.

It's also worth noting that while we derived the simplified model using a semi-classical approach, a fully classical model can be used as well.  The fully classical model is simple to derive, yields the same results, and can naturally be extended to a continuous rather than an impulse excitation.  Furthermore, the continuous excitation classical model results in precisely the full model in the low intensity limit given in Eq.~\ref{eq:Ip}.  We chose to use the semi-classical model in the manuscript because we feel it builds better intuition and, unlike the classical model, it can be easily extended to atoms with a more complicated excited state structure.  Also, while the results of the continuous drive model are a better approximation to what actually happens in the experiment, the resulting equation is more complicated (and, we feel, less intuitively useful) than the Lorentzian impulse excitation result.  For data analysis, we used the full quantum model --- the simplified model was presented only as a way to build intuition and to make quick approximate calculations.

To derive the fully classical model, one can simply note that the $m=-1$ and $m=1$ upper states are analogous to the two circular modes of a pendulum free to move in two dimensions.  Zeeman shifts can be incorporated by going to a reference frame which rotates at the Larmor frequency.  For an impulse drive, we simply assume that the pendulum is initially displaced in some direction.  For a continuous drive, one simply solves for a driven damped pendulum in which a sinusoidal driving force is applied in a direction which is constant in the rotating frame.  This is simple to do if the problem is solved in the non-rotating frame (with a driving force whose direction rotates at the Larmor frequency), and then the results are transformed back into the rotating frame.


\begin{thebibliography}{30}
\expandafter\ifx\csname natexlab\endcsname\relax\def\natexlab#1{#1}\fi
\expandafter\ifx\csname bibnamefont\endcsname\relax
  \def\bibnamefont#1{#1}\fi
\expandafter\ifx\csname bibfnamefont\endcsname\relax
  \def\bibfnamefont#1{#1}\fi
\expandafter\ifx\csname citenamefont\endcsname\relax
  \def\citenamefont#1{#1}\fi
\expandafter\ifx\csname url\endcsname\relax
  \def\url#1{\texttt{#1}}\fi
\expandafter\ifx\csname urlprefix\endcsname\relax\def\urlprefix{URL }\fi
\providecommand{\bibinfo}[2]{#2}
\providecommand{\eprint}[2][]{\url{#2}}

\bibitem[{\citenamefont{Moruzzi and Strumia}(2013)}]{moruzzi2013hanle}
\bibinfo{author}{\bibfnamefont{G.}~\bibnamefont{Moruzzi}} \bibnamefont{and}
  \bibinfo{author}{\bibfnamefont{F.}~\bibnamefont{Strumia}},
  \emph{\bibinfo{title}{The {H}anle effect and level-crossing spectroscopy}}
  (\bibinfo{publisher}{Springer Science \& Business Media},
  \bibinfo{year}{2013}).

\bibitem[{\citenamefont{Kastler}(1973)}]{kastler73he}
\bibinfo{author}{\bibfnamefont{A.}~\bibnamefont{Kastler}},
  \bibinfo{journal}{Nuclear Instruments and Methods}
  \textbf{\bibinfo{volume}{110}}, \bibinfo{pages}{259} (\bibinfo{year}{1973}).

\bibitem[{\citenamefont{Roux et~al.}(2008)\citenamefont{Roux, Emmert, Lupascu,
  Nirrengarten, Nogues, Brune, Raimond, and Haroche}}]{roux2008bose}
\bibinfo{author}{\bibfnamefont{C.}~\bibnamefont{Roux}},
  \bibinfo{author}{\bibfnamefont{A.}~\bibnamefont{Emmert}},
  \bibinfo{author}{\bibfnamefont{A.}~\bibnamefont{Lupascu}},
  \bibinfo{author}{\bibfnamefont{T.}~\bibnamefont{Nirrengarten}},
  \bibinfo{author}{\bibfnamefont{G.}~\bibnamefont{Nogues}},
  \bibinfo{author}{\bibfnamefont{M.}~\bibnamefont{Brune}},
  \bibinfo{author}{\bibfnamefont{J.-M.} \bibnamefont{Raimond}},
  \bibnamefont{and} \bibinfo{author}{\bibfnamefont{S.}~\bibnamefont{Haroche}},
  \bibinfo{journal}{Europhysics Letters} \textbf{\bibinfo{volume}{81}},
  \bibinfo{pages}{56004} (\bibinfo{year}{2008}).

\bibitem[{\citenamefont{Budker et~al.}(2002)\citenamefont{Budker, Gawlik,
  Kimball, Rochester, Yashchuk, and Weis}}]{budker02review}
\bibinfo{author}{\bibfnamefont{D.}~\bibnamefont{Budker}},
  \bibinfo{author}{\bibfnamefont{W.}~\bibnamefont{Gawlik}},
  \bibinfo{author}{\bibfnamefont{D.~F.} \bibnamefont{Kimball}},
  \bibinfo{author}{\bibfnamefont{S.~M.} \bibnamefont{Rochester}},
  \bibinfo{author}{\bibfnamefont{V.~V.} \bibnamefont{Yashchuk}},
  \bibnamefont{and} \bibinfo{author}{\bibfnamefont{A.}~\bibnamefont{Weis}},
  \bibinfo{journal}{Reviews of Modern Physics} \textbf{\bibinfo{volume}{74}},
  \bibinfo{pages}{1153} (\bibinfo{year}{2002}).

\bibitem[{\citenamefont{Johnson and Silsbee}(1985)}]{Johnson85ic}
\bibinfo{author}{\bibfnamefont{M.}~\bibnamefont{Johnson}} \bibnamefont{and}
  \bibinfo{author}{\bibfnamefont{R.~H.} \bibnamefont{Silsbee}},
  \bibinfo{journal}{Physical Review Letters} \textbf{\bibinfo{volume}{55}},
  \bibinfo{pages}{1790} (\bibinfo{year}{1985}).

\bibitem[{\citenamefont{O'Brien et~al.}(2016)\citenamefont{O'Brien, Spivak,
  Krueger, Peterson, Erickson, Bolon, Geppert, Leighton, and
  Crowell}}]{OBrien16oa}
\bibinfo{author}{\bibfnamefont{L.}~\bibnamefont{O'Brien}},
  \bibinfo{author}{\bibfnamefont{D.}~\bibnamefont{Spivak}},
  \bibinfo{author}{\bibfnamefont{N.}~\bibnamefont{Krueger}},
  \bibinfo{author}{\bibfnamefont{T.~A.} \bibnamefont{Peterson}},
  \bibinfo{author}{\bibfnamefont{M.~J.} \bibnamefont{Erickson}},
  \bibinfo{author}{\bibfnamefont{B.}~\bibnamefont{Bolon}},
  \bibinfo{author}{\bibfnamefont{C.~C.} \bibnamefont{Geppert}},
  \bibinfo{author}{\bibfnamefont{C.}~\bibnamefont{Leighton}}, \bibnamefont{and}
  \bibinfo{author}{\bibfnamefont{P.~A.} \bibnamefont{Crowell}},
  \bibinfo{journal}{Phys. Rev. B} \textbf{\bibinfo{volume}{94}},
  \bibinfo{pages}{094431} (\bibinfo{year}{2016}).

\bibitem[{\citenamefont{Colegrove et~al.}(1959)\citenamefont{Colegrove,
  Franken, Lewis, and Sands}}]{colegrove59levelcrossing}
\bibinfo{author}{\bibfnamefont{F.~D.} \bibnamefont{Colegrove}},
  \bibinfo{author}{\bibfnamefont{P.~A.} \bibnamefont{Franken}},
  \bibinfo{author}{\bibfnamefont{R.~R.} \bibnamefont{Lewis}}, \bibnamefont{and}
  \bibinfo{author}{\bibfnamefont{R.~H.} \bibnamefont{Sands}},
  \bibinfo{journal}{Physical Review Le} \textbf{\bibinfo{volume}{3}},
  \bibinfo{pages}{420} (\bibinfo{year}{1959}).

\bibitem[{\citenamefont{Novikov et~al.}(1974)\citenamefont{Novikov, Skrotskii,
  and Solomakh}}]{novikov74HanleEffect}
\bibinfo{author}{\bibfnamefont{L.~N.} \bibnamefont{Novikov}},
  \bibinfo{author}{\bibfnamefont{G.~V.} \bibnamefont{Skrotskii}},
  \bibnamefont{and} \bibinfo{author}{\bibfnamefont{G.~I.}
  \bibnamefont{Solomakh}}, \bibinfo{journal}{Usp. Fiz. Nauk}
  \textbf{\bibinfo{volume}{113}}, \bibinfo{pages}{597} (\bibinfo{year}{1974}).

\bibitem[{\citenamefont{Kaiser et~al.}(1991)\citenamefont{Kaiser,
  Vansteenkiste, Aspect, Arimondo, and Cohen-Tannoudji}}]{kaiser1991mechanical}
\bibinfo{author}{\bibfnamefont{R.}~\bibnamefont{Kaiser}},
  \bibinfo{author}{\bibfnamefont{N.}~\bibnamefont{Vansteenkiste}},
  \bibinfo{author}{\bibfnamefont{A.}~\bibnamefont{Aspect}},
  \bibinfo{author}{\bibfnamefont{E.}~\bibnamefont{Arimondo}}, \bibnamefont{and}
  \bibinfo{author}{\bibfnamefont{C.}~\bibnamefont{Cohen-Tannoudji}},
  \bibinfo{journal}{Zeitschrift f\"ur Physik D Atoms, Molecules and Clusters}
  \textbf{\bibinfo{volume}{18}}, \bibinfo{pages}{17} (\bibinfo{year}{1991}),
  ISSN \bibinfo{issn}{0178-7683},
  \urlprefix\url{http://dx.doi.org/10.1007/BF01418164}.

\bibitem[{\citenamefont{Rasmussen et~al.}(1974)\citenamefont{Rasmussen,
  Schieder, and Walther}}]{WaltherAtomicBeam}
\bibinfo{author}{\bibfnamefont{W.}~\bibnamefont{Rasmussen}},
  \bibinfo{author}{\bibfnamefont{R.}~\bibnamefont{Schieder}}, \bibnamefont{and}
  \bibinfo{author}{\bibfnamefont{H.}~\bibnamefont{Walther}},
  \bibinfo{journal}{Optical Communications} \textbf{\bibinfo{volume}{12}},
  \bibinfo{pages}{315} (\bibinfo{year}{1974}).

\bibitem[{\citenamefont{Avan and Cohen-Tannoudji}(1975)}]{ctmastereq}
\bibinfo{author}{\bibfnamefont{P.}~\bibnamefont{Avan}} \bibnamefont{and}
  \bibinfo{author}{\bibfnamefont{C.}~\bibnamefont{Cohen-Tannoudji}},
  \bibinfo{journal}{Le Journal de Physique - Letters}
  \textbf{\bibinfo{volume}{36}}, \bibinfo{pages}{L85} (\bibinfo{year}{1975}).

\bibitem[{\citenamefont{Franken}(1961)}]{franken61crossedstates}
\bibinfo{author}{\bibfnamefont{P.~A.} \bibnamefont{Franken}},
  \bibinfo{journal}{Physical Review} \textbf{\bibinfo{volume}{121}},
  \bibinfo{pages}{508} (\bibinfo{year}{1961}).

\bibitem[{\citenamefont{Stenflo}(1982)}]{stenflo1982hanle}
\bibinfo{author}{\bibfnamefont{J.~O.} \bibnamefont{Stenflo}},
  \bibinfo{journal}{Solar Physics} \textbf{\bibinfo{volume}{80}},
  \bibinfo{pages}{209} (\bibinfo{year}{1982}).

\bibitem[{\citenamefont{Stenflo et~al.}(1998)\citenamefont{Stenflo, Keller, and
  Gandorfer}}]{stenflo1998hanle}
\bibinfo{author}{\bibfnamefont{J.~O.} \bibnamefont{Stenflo}},
  \bibinfo{author}{\bibfnamefont{C.~U.} \bibnamefont{Keller}},
  \bibnamefont{and}
  \bibinfo{author}{\bibfnamefont{A.}~\bibnamefont{Gandorfer}},
  \bibinfo{journal}{Astronomy and Astrophysics} \textbf{\bibinfo{volume}{329}},
  \bibinfo{pages}{319} (\bibinfo{year}{1998}).

\bibitem[{\citenamefont{Bommier}(1980)}]{bommier1980quantum}
\bibinfo{author}{\bibfnamefont{V.}~\bibnamefont{Bommier}},
  \bibinfo{journal}{Astronomy and Astrophysics} \textbf{\bibinfo{volume}{87}},
  \bibinfo{pages}{109} (\bibinfo{year}{1980}).

\bibitem[{\citenamefont{Presura}(2012)}]{presura2012hanle}
\bibinfo{author}{\bibfnamefont{R.}~\bibnamefont{Presura}},
  \bibinfo{journal}{Review of Scientific Instruments}
  \textbf{\bibinfo{volume}{83}}, \bibinfo{eid}{10D528} (\bibinfo{year}{2012}),
  \urlprefix\url{http://scitation.aip.org/content/aip/journal/rsi/83/10/10.1063/1.4739233}.

\bibitem[{\citenamefont{Ignace et~al.}(1997)\citenamefont{Ignace, Nordsieck,
  and Cassinelli}}]{Ignace97he}
\bibinfo{author}{\bibfnamefont{R.}~\bibnamefont{Ignace}},
  \bibinfo{author}{\bibfnamefont{K.~H.} \bibnamefont{Nordsieck}},
  \bibnamefont{and} \bibinfo{author}{\bibfnamefont{J.~P.}
  \bibnamefont{Cassinelli}}, \bibinfo{journal}{The Astrophysical Journal}
  \textbf{\bibinfo{volume}{486}}, \bibinfo{pages}{550} (\bibinfo{year}{1997}).

\bibitem[{\citenamefont{Ignace}(2010)}]{Ignace10he}
\bibinfo{author}{\bibfnamefont{R.}~\bibnamefont{Ignace}}, \bibinfo{journal}{The
  Astro} \textbf{\bibinfo{volume}{725}}, \bibinfo{pages}{1040}
  (\bibinfo{year}{2010}).

\bibitem[{\citenamefont{Courteille et~al.}(2001)\citenamefont{Courteille,
  Muniz, Magalh\~aes, Kaiser, Marcassa, and
  Bagnato}}]{courteille2001tomography}
\bibinfo{author}{\bibfnamefont{P.}~\bibnamefont{Courteille}},
  \bibinfo{author}{\bibfnamefont{S.}~\bibnamefont{Muniz}},
  \bibinfo{author}{\bibfnamefont{K.}~\bibnamefont{Magalh\~aes}},
  \bibinfo{author}{\bibfnamefont{R.}~\bibnamefont{Kaiser}},
  \bibinfo{author}{\bibfnamefont{L.}~\bibnamefont{Marcassa}}, \bibnamefont{and}
  \bibinfo{author}{\bibfnamefont{V.}~\bibnamefont{Bagnato}},
  \bibinfo{journal}{The European Physical Journal D - Atomic, Molecular,
  Optical and Plasma Physics} \textbf{\bibinfo{volume}{15}},
  \bibinfo{pages}{173} (\bibinfo{year}{2001}), ISSN \bibinfo{issn}{1434-6060},
  \urlprefix\url{http://dx.doi.org/10.1007/s100530170163}.

\bibitem[{\citenamefont{Nielsen et~al.}(2008)\citenamefont{Nielsen, Lyngs\o,
  Thorseth, Galouzis, Therkildsen, van Ooijen, and
  Thomsen}}]{nielsen2008zeeman}
\bibinfo{author}{\bibfnamefont{C.~V.} \bibnamefont{Nielsen}},
  \bibinfo{author}{\bibfnamefont{J.~K.} \bibnamefont{Lyngs\o}},
  \bibinfo{author}{\bibfnamefont{A.}~\bibnamefont{Thorseth}},
  \bibinfo{author}{\bibfnamefont{M.}~\bibnamefont{Galouzis}},
  \bibinfo{author}{\bibfnamefont{K.~T.} \bibnamefont{Therkildsen}},
  \bibinfo{author}{\bibfnamefont{E.~D.} \bibnamefont{van Ooijen}},
  \bibnamefont{and} \bibinfo{author}{\bibfnamefont{J.~W.}
  \bibnamefont{Thomsen}}, \bibinfo{journal}{The European Physical Journal D}
  \textbf{\bibinfo{volume}{48}}, \bibinfo{pages}{111} (\bibinfo{year}{2008}),
  ISSN \bibinfo{issn}{1434-6060},
  \urlprefix\url{http://dx.doi.org/10.1140/epjd/e2008-00068-4}.

\bibitem[{\citenamefont{Jackson}(2016)}]{jackson2016situ}
\bibinfo{author}{\bibfnamefont{J.~S.} \bibnamefont{Jackson}}, Master's thesis,
  \bibinfo{school}{Brigham Young University} (\bibinfo{year}{2016}),
  \urlprefix\url{http://scholarsarchive.byu.edu/etd/5982}.

\bibitem[{\citenamefont{Erickson}(2011)}]{erickson2011development}
\bibinfo{author}{\bibfnamefont{C.~J.} \bibnamefont{Erickson}}, Ph.D. thesis,
  \bibinfo{school}{Brigham Young University} (\bibinfo{year}{2011}).

\bibitem[{vie()}]{viewingstereopairs}
\bibinfo{note}{Https://en.wikipedia.org/wiki/Stereoscopy\#Freeviewing}.

\bibitem[{\citenamefont{Renzoni et~al.}(2001)\citenamefont{Renzoni, Zimmermann,
  Verkerk, and Arimondo}}]{Arimondo01HanleClosed}
\bibinfo{author}{\bibfnamefont{F.}~\bibnamefont{Renzoni}},
  \bibinfo{author}{\bibfnamefont{C.}~\bibnamefont{Zimmermann}},
  \bibinfo{author}{\bibfnamefont{P.}~\bibnamefont{Verkerk}}, \bibnamefont{and}
  \bibinfo{author}{\bibfnamefont{E.}~\bibnamefont{Arimondo}},
  \bibinfo{journal}{Journal of Optics B:Quantum and Semiclassical Optics}
  \textbf{\bibinfo{volume}{3}}, \bibinfo{pages}{S7} (\bibinfo{year}{2001}).

\bibitem[{\citenamefont{Grewal and
  Pattabiraman}(2015)}]{Singh15GroundStateHanleInRb}
\bibinfo{author}{\bibfnamefont{R.~S.} \bibnamefont{Grewal}} \bibnamefont{and}
  \bibinfo{author}{\bibfnamefont{M.}~\bibnamefont{Pattabiraman}},
  \bibinfo{journal}{Journal of Physics B: Atomic, Molecular and Optical
  Physics} \textbf{\bibinfo{volume}{48}}, \bibinfo{pages}{085501}
  (\bibinfo{year}{2015}).

\bibitem[{\citenamefont{Arimondo}(2007)}]{arimondo07DarkResonances}
\bibinfo{author}{\bibfnamefont{E.}~\bibnamefont{Arimondo}}, in
  \emph{\bibinfo{booktitle}{Proceedings of the International School and
  Conference on Optics and Optical Materials}} (\bibinfo{year}{2007}).

\bibitem[{\citenamefont{Budker et~al.}(2008)\citenamefont{Budker, Kimball, and
  De{M}ille}}]{budkeratomicphysics}
\bibinfo{author}{\bibfnamefont{D.}~\bibnamefont{Budker}},
  \bibinfo{author}{\bibfnamefont{D.~F.} \bibnamefont{Kimball}},
  \bibnamefont{and} \bibinfo{author}{\bibfnamefont{D.~P.}
  \bibnamefont{De{M}ille}}, \emph{\bibinfo{title}{Atomic Physics}}
  (\bibinfo{publisher}{Oxford University Press}, \bibinfo{year}{2008}),
  \bibinfo{edition}{second edition} ed.

\bibitem[{mag()}]{magnetstrengths}
\bibinfo{note}{Https://www.supermagnete.de/eng/data\_table.php},
  \urlprefix\url{https://www.supermagnete.de/eng/data_table.php}.

\bibitem[{\citenamefont{Richards et~al.}(2013)\citenamefont{Richards, Kazmi,
  Davis, Olin, and Dunn}}]{richards2013logitech}
\bibinfo{author}{\bibfnamefont{L.~M.} \bibnamefont{Richards}},
  \bibinfo{author}{\bibfnamefont{S.~M.~S.} \bibnamefont{Kazmi}},
  \bibinfo{author}{\bibfnamefont{J.~L.} \bibnamefont{Davis}},
  \bibinfo{author}{\bibfnamefont{K.~E.} \bibnamefont{Olin}}, \bibnamefont{and}
  \bibinfo{author}{\bibfnamefont{A.~K.} \bibnamefont{Dunn}},
  \bibinfo{journal}{Biomed. Opt. Express} \textbf{\bibinfo{volume}{4}},
  \bibinfo{pages}{2269} (\bibinfo{year}{2013}),
  \urlprefix\url{http://www.osapublishing.org/boe/abstract.cfm?URI=boe-4-10-2269}.

\bibitem[{\citenamefont{Langer et~al.}(2013)\citenamefont{Langer, Hochreiner,
  Burgholzer, and Berer}}]{langer2013logitech}
\bibinfo{author}{\bibfnamefont{G.}~\bibnamefont{Langer}},
  \bibinfo{author}{\bibfnamefont{A.}~\bibnamefont{Hochreiner}},
  \bibinfo{author}{\bibfnamefont{P.}~\bibnamefont{Burgholzer}},
  \bibnamefont{and} \bibinfo{author}{\bibfnamefont{T.}~\bibnamefont{Berer}},
  \bibinfo{journal}{Optics and Lasers in Engineering}
  \textbf{\bibinfo{volume}{51}}, \bibinfo{pages}{571 } (\bibinfo{year}{2013}),
  ISSN \bibinfo{issn}{0143-8166},
  \urlprefix\url{http://www.sciencedirect.com/science/article/pii/S014381661200348X}.

\end{thebibliography}

\section*{Acknowledgements}

We gratefully acknowledge support from NSF grant No. PHY1205736.

\section*{Author contributions statement}

Both authors conceived the methods and experiments.  J.J. conducted the experiments with consultation from D.D.  Both authors analyzed the results.  Both authors reviewed the manuscript.

\section*{Competing interests}
The authors declare no competing interests.

\end{document}